**Title:** **Mechanics of the cellular actin cortex: from signalling to shape change**

Manasi Kelkar[1,*], Pierre Bohec[1,*], Guillaume Charras[1,2,3]

[1]London Centre for Nanotechnology, University College London, London, WC1H 0AH, UK
[2]Department of Cell and Developmental Biology, University College London, London, WC1E 6BT, UK
[3]Institute for the Physics of Living Systems, University College London, London, WC1E 6BT, UK

*Equal contribution

Author for correspondence: Guillaume Charras, g.charras@ucl.ac.uk

**Abstract:**
The actin cortex is a thin layer of actin, myosin, and actin binding proteins that underlies the membrane of most animal cells. It is highly dynamic and can undergo remodelling on time-scales of tens of seconds thanks to protein turnover and myosin-mediated contractions. The cortex enables cells to resist external mechanical stresses, controls cell shape, and allows cells to exert forces on their neighbours. Thus, its mechanical properties are key to its physiological function. Here, we give an overview of how cortex composition, structure, and dynamics control cortex mechanics and cell shape. We use mitosis as an example to illustrate how global and local regulation of cortex mechanics give rise to a complex series of cell shape changes.

## Introduction

The cortex is a dense submembranous network of actin filaments, actin binding proteins and myosin about 200nm thick. The turnover of cortical proteins together with the contractile forces exerted by myosin motors endow the cortex with complex mechanical behaviours and give rise to a cellular-scale surface tension. Inhomogeneities in cortical tension drive the shape changes that occur during cell division, migration and tissue morphogenesis. While our understanding of the signalling and cytoskeletal machinery governing morphogenesis is extensive, we know less about the changes in cortical organisation and mechanics that drive shape change.

A clear illustration of the importance of cortex mechanics for cellular shape changes comes from cell division. During mitosis, a global increase in myosin activity leads to a threefold increase in cortical stiffness that drives cell rounding in metaphase [1-3]. Later, local changes in cortex mechanics give rise to tension gradients that cause cell deformations. In cytokinesis, cortical stiffness is highest at the equator [4], reflective of a locally higher tension necessary to drive furrowing [5]. Asymmetric regulation of polar acto-myosin contractility produces unequally sized daughter cells in *C.elegans* neuroblasts [6] and misregulation results in de-stabilisation of the cytokinetic furrow [7]. In addition to localized changes in cell shape, contractility gradients can lead to cortical flows in the plane of the membrane. In cytokinesis, actomyosin flows from the poles to the equator contribute to cleavage furrow formation [8,9] and modelling suggests that they could contribute to orientation of actin filaments into a ring-like structure [10].

In this brief review, we first give an overview of the mechanical properties of the cortex before examining how changes in organisation and contractility affect cortical mechanics. We then examine the signalling pathways that control cortical mechanics on a global and local scale focusing on mitotic shape changes.

## Actomyosin cortex mechanics

The basic principle of most mechanical characterisation techniques is to apply a defined force and measure the resulting deformation (for a solid-like material) and/or deformation speed (for a fluid-like material) (**Fig 1A**). To enable comparison across objects with different geometries, these responses are normalised to obtain **stress-strain** or **stress-strain rate** relationships respectively defining the **elastic modulus** for a solid-like material and **viscosity** for a liquid-like material (**Figure 1B**). However, living objects are often neither completely liquid nor solid [11]. When we apply a force on the cortex, it will store part of the energy by increasing its internal stress, as solids do, and dissipate part of it by flowing, as a viscous fluid. From a physics perspective, the actomyosin cortex is a self-organized active polymer network (**Figure 2D-E**) that can be classified as an **active viscoelastic material or active gel** [12].

The cortex mechanical behaviour depends on the rate at which stress is applied and the time elapsed since loading. If the load is applied slowly, little elastic stress is stored and the cortex just flows. Whereas, when a load is applied rapidly, the cortex does not have time to flow and it stores stress at short time-scales before flowing at longer time-scales because of the reorganisation of the cortical network and the continuous turnover of its constituents [13,14]. Turnover occurs in seconds for crosslinkers [15,16], tens of seconds for actin filaments and individual myosins [9,17], and hundreds of seconds for myosin mini-filaments [18].

Cells were initially described as linear viscoelastic materials using equivalent mechanical circuits of connected elastic springs and viscous dashpots (**Figure 1B**). However, precise characterisation of the **local** and **global** viscoelastic properties of the cortex of spread and round cells (using for example optical and magnetic tweezers atomic force microscopy or micropipette aspiration [19], **Figure 1C**) revealed that cells [20] and the cortex possess a weak power law rheology [11], in between a liquid and a solid (**Figure 1B**). These rheological properties cannot easily be mimicked by linear viscoelasticity but new rheological elements based on fractional rheology (springpots) allow to capture power-law behaviours while retaining the conceptual simplicity of equivalent mechanical circuits [21].

Interestingly, many other materials such as foams, emulsions or pastes display the same non-linear flow behaviour. These materials are collectively referred to as presenting soft glassy rheology. What sets the cortex apart from inert soft glassy materials is the presence of active processes due to myosin **molecular motors** that transform chemical energy from ATP into fluctuating forces exerted on the actin filaments [22]. The active stress generated by molecular motors is often referred to as contractility, active stress or pre-stress and plays a central role in setting cellular mechanical properties [11]. Indeed, one needs only palpate a contracted muscle to realise how much contractility influences our perception of stiffness. Although the role of molecular motors has been intensely studied, other active processes such **actin polymerisation** may play a role.

As the cortex is very thin compared to the cell diameter, one can approximate the cell to a liquid droplet and the cortex to a two-dimensional shell under internal stress [23] (**Figure 2A, B**). By integrating internal stress across cortical thickness [24], one obtains a cortical tension expressed as force per unit length (**Figure 2A,C**) [25] that can be characterised experimentally [23,26,27]. Cortical tension comprises a contribution from active stress, called **active tension** arising from motor activity**,** and a passive contribution due to the viscoelastic nature of the cortex that arises in response to flow or deformation [14]. In summary, a minimal mechanical description of the cortex comprises a cortical tension, a storage modulus (elasticity) and a loss modulus (viscosity).

## Regulation of cortical tension through actin network organisation

One current challenge is to understand how cortical mechanical properties emerge from the arrangement, abundance, activity, and biochemical properties of its constituents (**Figure 2D-E**). From gel physics, we predict that the length, the degree of crosslinking, and alignment of actin filaments will influence cortical network mechanics; while from biology, we expect that the amount and activity of molecular motors in the cortex will govern cortical tension. Here, we first examine how regulating the F-actin network affects cortical mechanics before examining the role of myosin and signalling in the following section.

**Actin nucleation**
The genesis of the cortex relies on *de novo* nucleation of actin filaments [28] with two main nucleators identified: the Arp2/3 complex that catalyses formation of branched actin networks and the formin mDia1 that polymerises linear filament arrays (**Figure 3A**). These are also present in the cortex of mitotic cells [28] and, although only mDia1 depletion compromises cell division, co-inhibition of Arp2/3 potentiates this effect suggesting synergistic interactions between the two actin nucleators [28]. Interplay between formins and Arp2/3 in the cortex also appears in invertebrates such as *Drosophila* [29] and *C.elegans* [30]. In *Dictyostelium*, the mDia1 related formins (ForA, ForE and ForH) act together to maintain cortical integrity during development and cytokinesis [31].

**Actin filament length**
Computational simulations and experiments indicate that a change in average filament length alone may be sufficient to modulate cortical tension and that tension possesses a length optimum [27] (**Figure 3C**). Indeed, depletion of proteins that regulate filament length (either through depolymerisation or polymerisation) leads to a decrease in tension. The existence of such an optimum may be because precise regulation of filament length and tension is crucial to allow error-free division. Indeed, depletion of either the actin severing protein WDR1/Aip1 [32], the dis-assembly protein ADF/cofilin [33] or the assembly protein profilin [34] all lead to cytokinetic defects.

One current limitation to computational approaches is our lack of knowledge of the length distribution of cortical actin filaments. Although several subpopulations of filaments with different dynamics have been identified in the cortex [17], characterising the filament length distribution remains very challenging. Indeed, scanning electron microscopy images of HeLa cells in mitosis reveal a mesh size on the order of 20-30nm [27] (**Figure 2D-E**), still beyond reach of super-resolution microscopy or atomic force microscopy which can only resolve less dense meshworks in specific cell types [35]. An alternative is single molecule imaging to track single formins or actin monomers with the assumption that their movement reflects filament barbed-end elongation and treadmilling respectively [36]. Future advances in microscopy will likely shed light on these issues.

**Actin network organization**
Intuitively, we realise that the degree of alignment of F-actin filaments within a network will influence tension generation by myosins. Indeed, myosins attaching to antiparallel filament arrays produce force more efficiently than those binding to isotropic networks like the cortex [37]. Therefore, network transitions from isotropic to ordered could give rise to dramatic changes in tension without necessitating additional myosin recruitment [10]. Such nematic transitions may be at play during division. Early reports using transmission electron microscopy indicated the presence of aligned antiparallel actin filaments in the cleavage furrow [38,39] and more recent findings have revealed a remarkable degree of alignment of myosin mini-filaments in this region [40,41]. During pseudocleavage formation in C elegans, cortical flows of actomyosin lead to dynamic ordering of filaments [42]. Altogether, there appears to be a transition in actin filament organisation from isotropic in metaphase to ordered in the furrow region (**Figure 3D**) which may lead to the large local increase in tension driving division.

As several different actin nucleators generate cortical F-actin, changing their relative activity may allow fine-tuning of network organisation. In addition to RhoGTPases, the activity of nucleators is modulated by nucleation promoting factors (NPFs), several of which are present in the cortex [43]. Recent work has shown that competition between two cortical NPFs, the Wave Regulatory Complex (WRC) and SPIN90, for Arp2/3 modulates the degree of branching in F-actin networks [43,44]. Furthermore, SPIN90 can form a ternary complex with Arp2/3 and mDia1 to increase generation of long filaments (**Figure 3B**). Therefore, the interplay of mDia1, Arp2/3, WRC and SPIN90 can modulate the degree of branching and the length of filaments allowing to switch between lamellipodial-like organisations dominated by Arp2/3 and WRC to cortical organisations in which all four proteins are active [43] (**Figure 3B**).

**Actin network connectivity**
In addition to myosin, which can act as a crosslinker, many other actin crosslinking proteins have been identified in the cortex [45,46]. *In* vitro studies have highlighted the crucial role of crosslinking in determining network contractility [37,47]. Taken together, what is becoming evident is that cortical tension is maximum for an intermediate level of connectivity. If the network is too loosely connected stresses cannot propagate, while if there is too much connectivity, the network is rigid and cannot actively remodel (**Figure 3C**). In support of these views, changes in filament length and crosslinker abundance perturb stress generation. Depletion experiments have revealed roles for several crosslinkers in stress generation and mitotic rounding in HeLa cells and in *C.elegans* [14,48,49]. Conversely, overexpression of alpha-actinin in NRK cells causes cytokinetic defects and failures [16]; whereas, intermediate levels of alpha-actinin and the scaffold protein anillin result in maximal speed of ring constriction in *S.pombe* [50] and in *C.elegans* [51] respectively.

Recent work suggests that cells may actively modulate network connectivity in the actin cortex during cell shape changes. In addition to changes in protein abundance, connectivity can also be modulated through changes in activity or turnover rate. For example, the actin capping protein is recruited in the final stages of cytokinesis, perhaps to reduce connectivity by reducing filament length and allow continued force generation [52]. In the cleavage furrow, myosin-2 accelerates the turnover of actin whereas over-expression of alpha-actinin slows it [9,16]. In Dictyostelium, differential turnover rates of motors and crosslinkers at the equator and poles appear crucial in determining cortical stiffness and the rate of furrow ingression [15].

**<u>Spatio temporal regulation of contractility by Rho signalling</u>**
Perhaps the best understood mechanism of cortical tension control is through regulation of the abundance and activity of cortical myosins downstream of RhoGTPase signalling. Rho-GTPases cycle between an active GTP-bound and an inactive GDP-bound state thanks to regulatory proteins: Guanine nucleotide Exchange Factors (GEFs) favour exchange of GDP for GTP and GTPase Activating Proteins (GAPs) catalyse GTP hydrolysis (**Figure 4A**). Amongst RhoGTPases, RhoA has the best characterised cortical activity although recent work indicates that interplay between RhoC and RhoD may also be important [53]. RhoA activates Rho-kinase (ROCK) which increases myosin activity by direct phosphorylation and inhibition of dephosphorylation. In addition, RhoA modulates actin network organisation through direct activation of formins (**Figure 4A**) and indirect inactivation of cofilin via Lim-kinase. Thus, RhoA modulates both the actin scaffold and the motors. Basal control of cortical contractility in interphase remains poorly understood. However, experiments in blebs indicate that contractile activity is downregulated by Rnd3 and p190RhoGAP [54] and activated by interaction between MyoGEF and RhoA [55]. In *C.elegans,* single molecule actomyosin dynamics has revealed the interplay between RhoA and GAPs to generate pulses of contraction [56].

Mitosis involves a series of shape changes starting with rounding in prometaphase, polar relaxation and elongation in anaphase and furrowing in cytokinesis [57] (**Figure 4B**). Each of these events requires a precise regulation of contractility and cortical tension at a global or a local scale [2,7]. RhoA is indispensable for mitosis, operating in turn at a global or a local level.

First, at mitotic onset, a global increase in cortical tension regulates rounding [2]. Following nuclear envelope breakdown, re-localization of RhoGEF-Ect2 from the nucleus to the cortex and a simultaneous decrease in p190rhoGAP activity drive cell rounding [29,58,59]. Later, cdk1 activation of both ROCK and p21-activated kinases (PAK) leads to an increase in myosin recruitment at the cortex [1].

At anaphase, cortical myosin distribution becomes heterogeneous with enhanced accumulation at the midzone leading to a tension gradient that drives actomyosin flows and potentially nematic ordering. RhoA enrichment in this region is orchestrated by interplay between two GEFs (Ect2, GEF-H1) and two GAPs (MP-GAP and MgcRacGAP) [60,61] (**Figure 4C**). Simultaneously, accumulation of both Ran-GTP and the phosphatase PP1/sds22 at the poles leads to anillin clearing and loss of actomyosin respectively, softening the polar cortex [3,62,63]. Later, both MyoGEF and MgcRacGAP accumulate at the furrow and finally during abscission the GEF LARG and p50-RhoGAP localise to the midbody [64-66]. Concomitant with these changes in myosin localisation, cortical stiffness increases dramatically in the furrow region [4].

## Concluding remarks and future perspectives

The last few years have seen considerable progress in our understanding of how composition and organization of the cortex interplay to control force generation. Despite this, how and when each of the diverse processes modulating cortical tension contribute to generating shape change remains unclear. One challenge for the future will be to link spatio-temporal changes in signalling to molecular level cortex reorganisation and finally mechanical changes at the cellular level.

Answering these questions will likely require control of signalling in space and time using optogenetics. Interestingly, recent studies have revealed that localized activation of RhoA by optogenetic targeting of a RhoGEF (LARG or Ect2) to the equator is sufficient to initiate formation of an actomyosin-rich furrow [67,68]. However, in both cases, ingression was not robust, suggesting the requirement for additional factors either temporally or spatially. A full mechanical understanding of mitotic morphogenesis will necessitate characterisation of the contribution of each mitotic GEF/GAP to cortical organisation and mechanics.

Another key unresolved question is the interplay between the actin cortex and other cytoskeletal components. While it is widely accepted that astral microtubules interact with the cortex during mitosis, the exact nature of that interplay remains somewhat mysterious with microtubules traversing the cortex to bind to Gai/LGN/NuMA/dynein at the membrane. Interestingly, two recent studies reveal a role for vimentin intermediate filaments in regulating actin cortex thickness and tension during cell division [46,69].

**Acknowledgements:**
The authors apologise to colleagues whose work we have been unable to cite for reasons of space. MK was supported by an SNSF early postdoc fellowship P2LAP3_164919. GC was funded by a European Research Council consolidator grant (CoG-647186). PB was supported by a CRUK multidisciplinary grant to GC. GC acknowledges support from an HFSP Young Investigator grant (RGY66/2013).

The authors declare no conflict of interest.

## Figures:

## Rheology

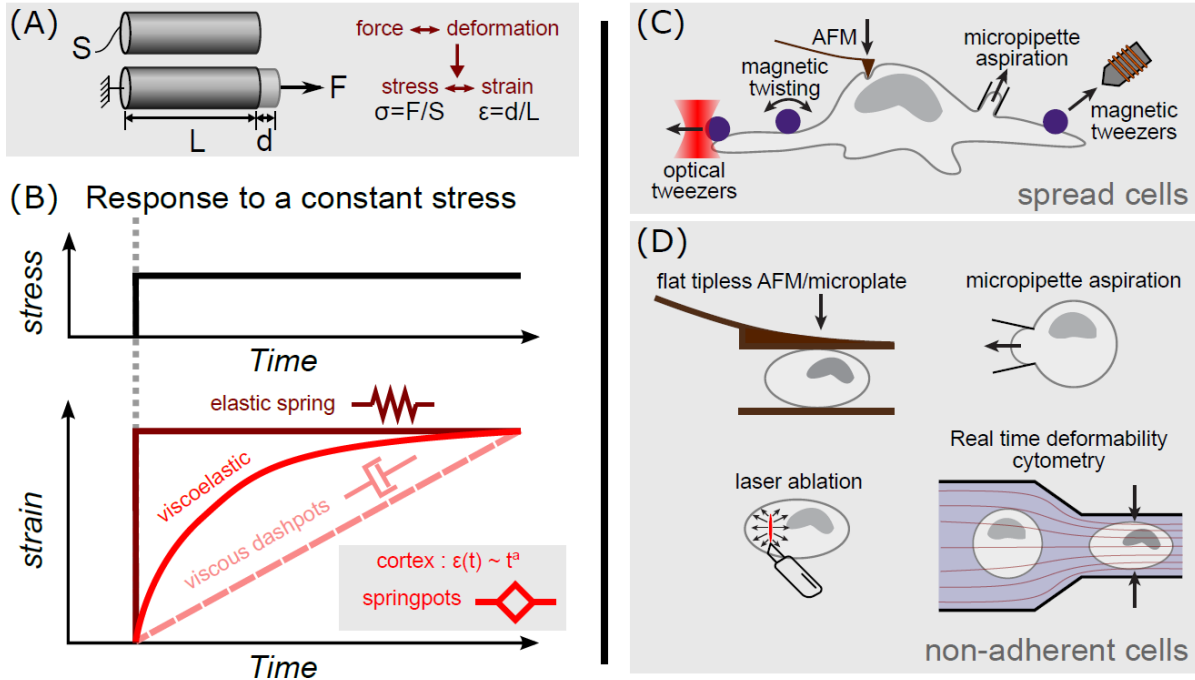

**Figure 1: Characterising cortical mechanics.** (**A**) Rheology is the study of the deformation (for solids) and the flow (for liquids) of materials in response to a force. A material of length L subjected at an external force F will deform by a length d. To enable comparison across objects with different geometries, the deformation d and force F are normalised by the object dimension to obtain the stress σ = F/S where S is the cross sectional area and the strain ε = d/L. (**B**) A typical rheology experiment measures the strain of an object subjected to a constant stress over time. From this experiment, one can define two limit cases. A constant stress on an elastic material, symbolized by a spring, will induce a constant deformation (burgundy curve). For a newtonian liquid, symbolized by a viscous dashpot, the strain is proportional to the application time of the stress (pink dashed curve). These behaviours reflect the fact that an elastic-like material will store all the energy while a liquid-like material will dissipate it by flowing. Beyond these two ideal cases, some materials can both store and dissipate the energy: the viscoelastic materials (red curves). Some of these have a linear viscoelastic behaviour while others (like the cortex) follow a scale-free power law rheology in which the strain evolves as ε(t) = $t^α$ with α~0.2. Materials with complex rheologies can be described with equivalent mechanical circuits comprising springs and dashpots (for linear viscoelasticity) or including springpots when they possess power law rheology. (**C**) Many techniques have been developed to study the local mechanical properties of the actomyosin cortex in spread cells. One strategy is to attach a probe to the actomyosin cortex via specific transmembrane proteins (e.g integrin). Then, a force can be exerted on this probe to deform the cortex either with optical tweezers or magnetic tweezers. Atomic force microscopy (AFM) uses a flexible cantilever with a tip to locally deform the cortex. A simple micropipette can also deform the cortex using suction. In these techniques, either the force is controlled and the deformation measured or the opposite. (**D**) Techniques for characterising cell-scale mechanical properties of the cortex operate on non-adherent cells. Microplates or tipless AFM cantilevers compress the cells between two plates, one of them being flexible. Micropipette aspiration can also be used on rounded cells. Laser ablation cuts the cortex and tracks its deformation to probe cortical tension.

Real-time deformability cytometry is a high-throughput method based on hydrodynamic deformation of the cells [70].

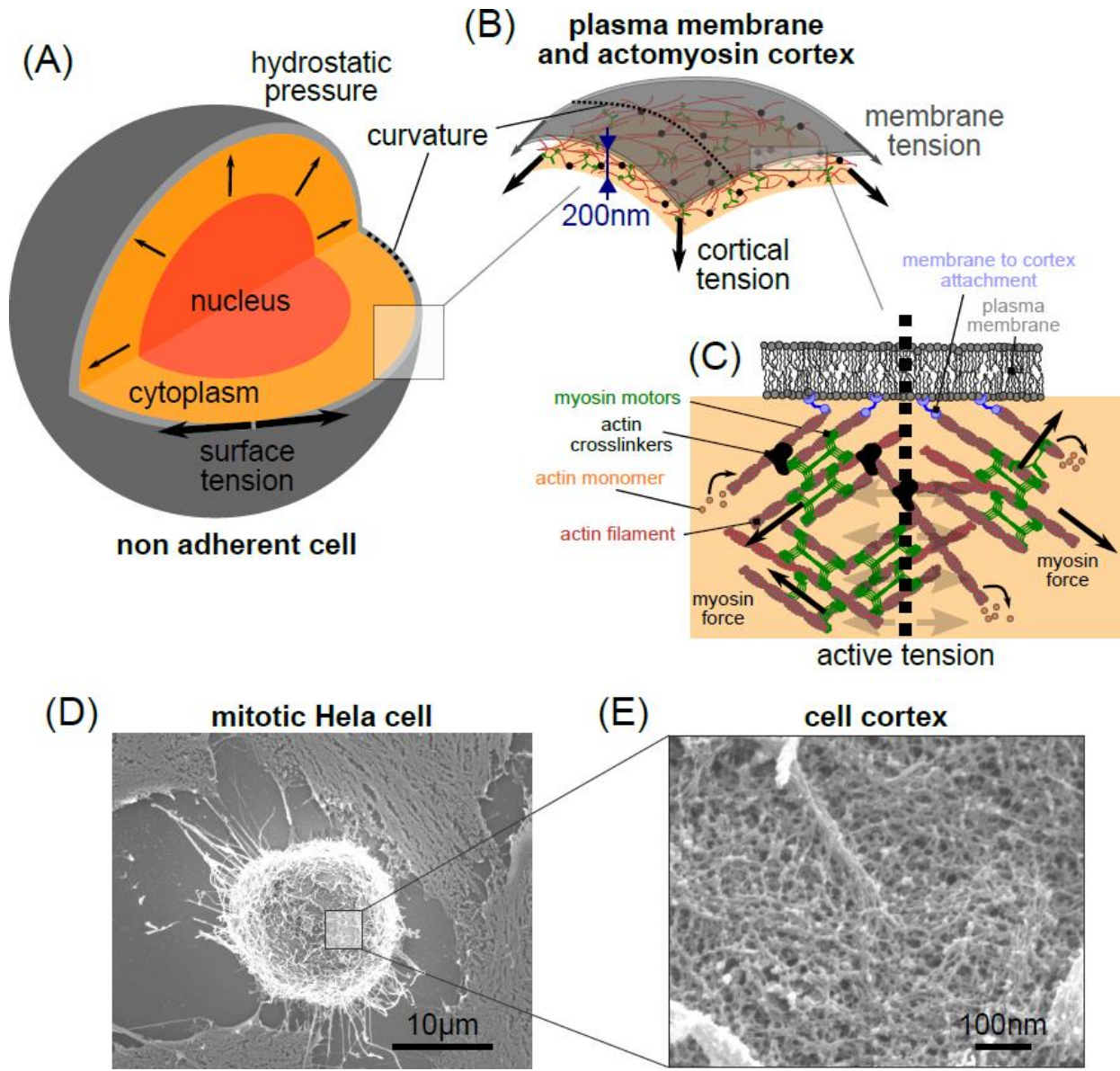

**Figure 2: Overview of the components and the mechanics of the actomyosin cortex**

(**A**) Diagram of a non-adherent or rounded cell. As the cortex is very thin (~200 nm) compared to the cell diameter (~10µm), one can approximate the cell to a liquid droplet surrounded by a two-dimensional shell. The surface tension of the cell is related to the cell curvature and the hydrostatic pressure in the liquid cell interior. (**B**) The cell shell is composed by the plasma membrane and the actomyosin cortex. The actomyosin cortex is coupled to the plasma membrane via linker proteins (e.g. FERM proteins, purple in C). Although both the plasma membrane and the actomyosin cortex contribute to the total surface tension of the cell, experimental work has shown that cortex tension dominates. (**C**) The cortex is composed of actin filaments (red), myosin motors (green) and crosslinkers (black) that dynamically turnover endowing it with complex rheological properties. (**D**) Electron micrograph of a mitotic Hela cell. The plasma membrane has been removed to reveal the structure of the cortex. A zoom of the boxed region is shown in E. (**E**) Electron micrograph of the actin cortex of a mitotic Hela cell. The mesh size of the cortical actin network is about 20nm.

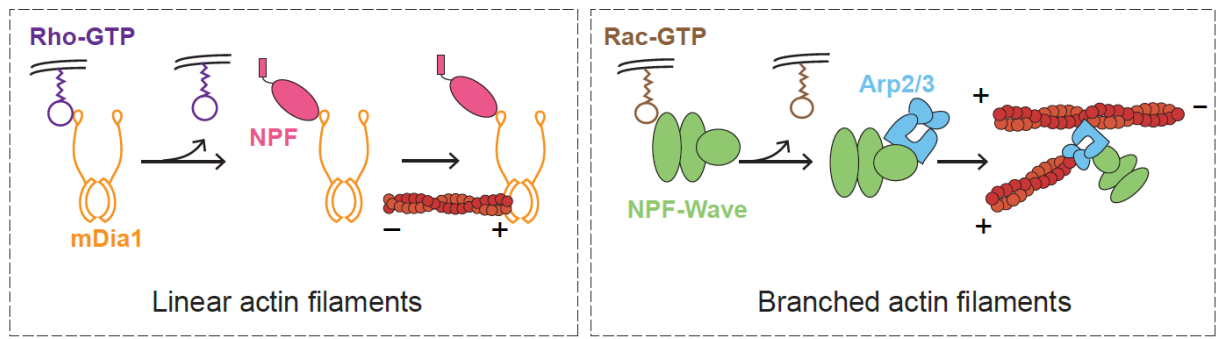

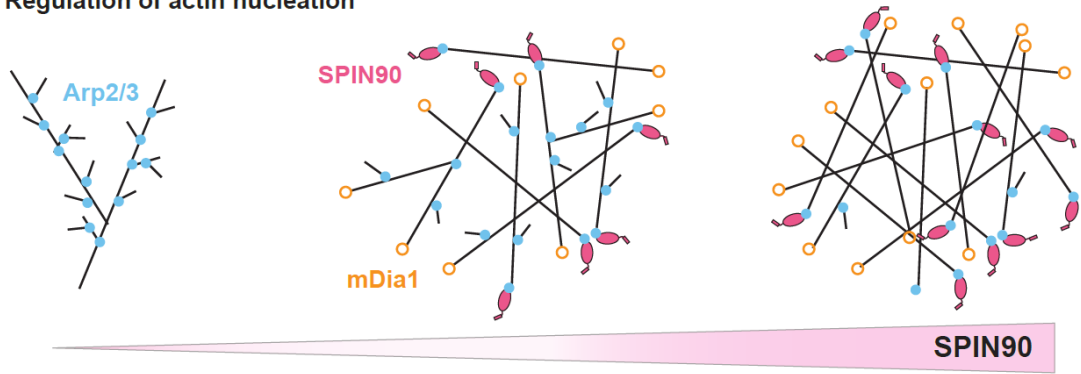

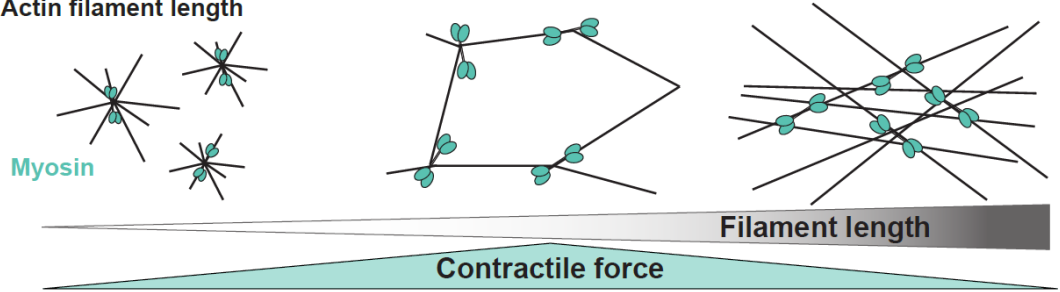

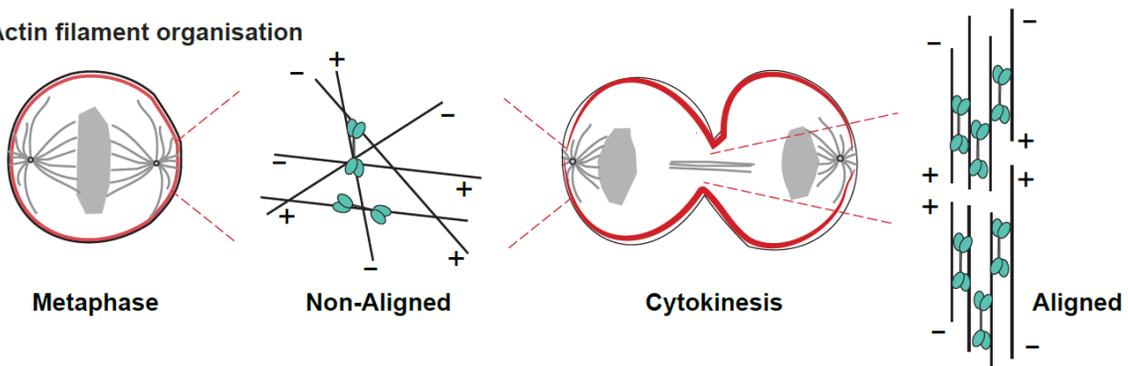

**Figure 3. Regulation of cortical tension through actin network organization** (A) Actin nucleation occurs by direct activation of formins by RhoA-GTP. These nucleate linear actin filaments, while the Arp2/3 complex nucleates branched networks. Arp2/3 is activated by Rac1-GTP indirectly via the Wave-complex. In addition the Rho-GTPases, both formins and Arp2/3 can be further regulated by nucleation promoting factors (NPFs) which either activate or maintain their activity. (**B**) Increasing the density of the NPF SPIN90 leads to lower branching densities and favours generation of linear filaments elongated by mDia1 thereby increasing filament length. (**C**) Intermediate length of filaments favour maximum tension generation. If filaments are too short, the network is loosely connected and, if they are too long, the network is too rigid to generate stress. (**D**) The efficiency of tension generation by myosins is modulated by network organisation. In metaphase, actin filaments in the cortex are isotropically oriented; whereas in the furrow region in anaphase, antiparallel actin filaments lead to increased stress generation.

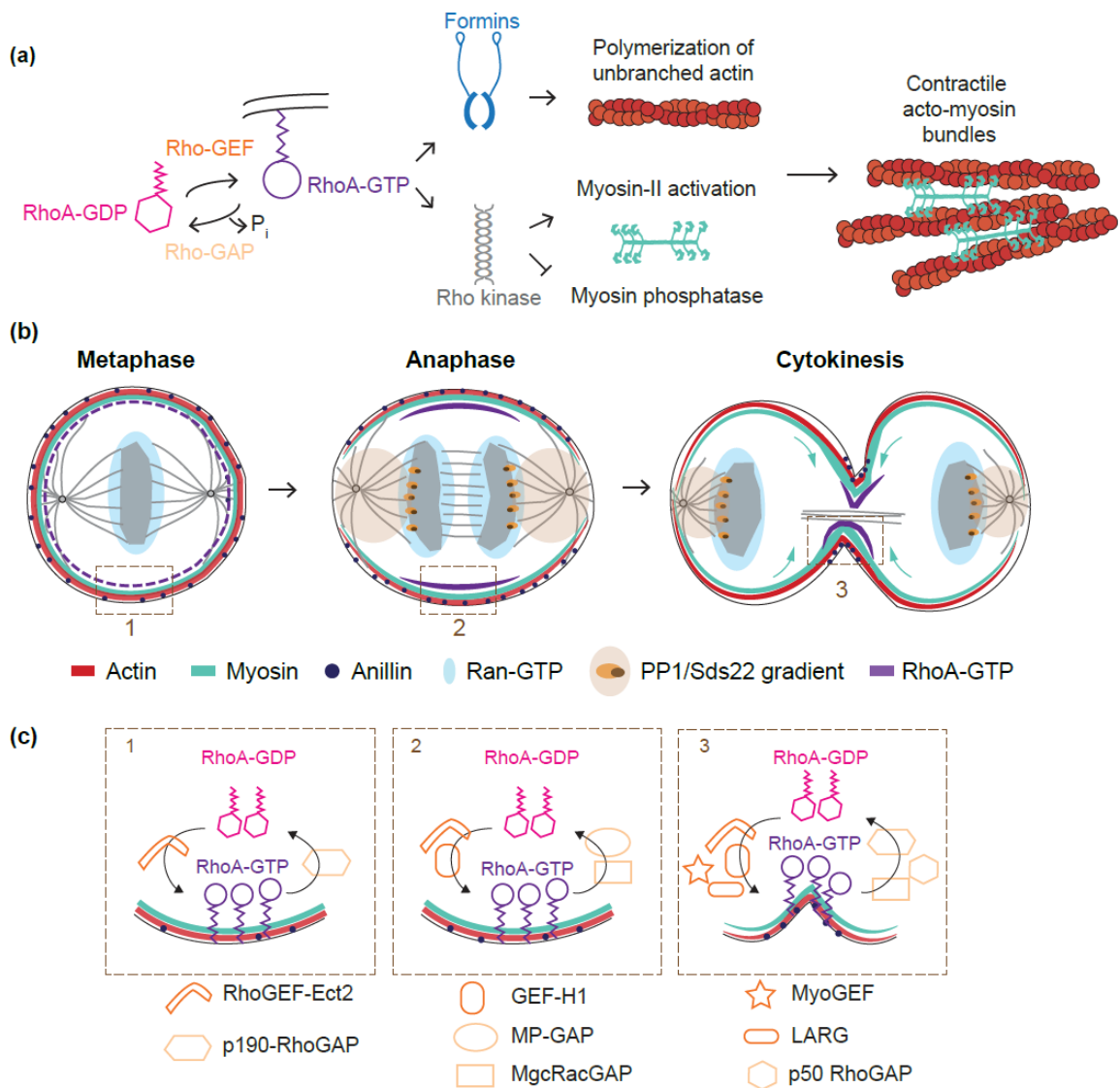

**Figure 4. Spatio temporal regulation of contractility by Rho signalling. (A)** Schematic of the RhoGTPase signalling cycle. RhoGEFs exchange GDP for GTP to activate RhoA, while RhoGAPs mediate GTP hydrolysis to inactivate RhoA. Active RhoA binds to formins and ROCK resulting in the formation of contractile actomyosin bundles. (**B**) Shape changes occurring in mitosis proceed from rounding in metaphase, elongation in anaphase to furrowing in cytokinesis. As cells enter mitosis, the actomyosin cortex (actin in red, myosin in green) becomes more contractile aiding cell rounding. Active RhoAGTP (purple) and the actin binding protein anillin (dark blue) are uniformly distributed along the cortex. At anaphase onset, a gradient of the phosphatase PP1-Sd22 (orange/brown) combined with anillin clearing relax contractility at the poles causing cell shape elongation. Simultaneous accumulation of actomyosin at the midzone occurs due to concerted action of multiple GEFs/GAPs. Finally, at cytokinesis RhoA signaling at the cleavage furrow aids in contractile ring assembly. Green arrows indicate actomyosin flows from the poles to the furrow region. (**C**) Schematic of the various GEFs and GAPs present at the cortex in metaphase, anaphase and cytokinesis.